\documentclass[letter]{aa}
\usepackage[varg]{txfonts}
\usepackage{graphicx}
\usepackage{natbib}
\usepackage{lipsum}
\bibpunct{(}{)}{;}{a}{}{,}

\usepackage[colorlinks=true,urlcolor=blue,citecolor=blue,linkcolor=blue]{hyperref}

\def\ha{H$\alpha$}

\def\CaII{\ion{Ca}{ii}}
\def\CaH{\CaII\, H}

\def\MgIIk{\ion{Mg}{ii}\,k}

\def\kms {{\mathrm{km}\,\mathrm{s}^{-1}}}

\begin{document}

\title{Chromospheric counterparts of solar transition region unresolved fine structure loops}
\titlerunning{{Chromospheric Counterparts to Transition Region Loops}}

\author{Tiago M. D. Pereira\inst{1, 2} \and Luc Rouppe van der Voort\inst{1, 2} \and Viggo H. Hansteen\inst{1, 2, 3} \and Bart De Pontieu\inst{3, 1, 2}}
\institute{Rosseland Centre for Solar Physics, University of Oslo, P.O. Box 1029 Blindern, NO--0315 Oslo, Norway\\ email: \texttt{tiago.pereira@astro.uio.no}
\and
Institute of Theoretical Astrophysics, University of Oslo, P.O. Box 1029 Blindern, NO--0315 Oslo, Norway
\and Lockheed Martin Solar and Astrophysics Laboratory, Lockheed Martin Advanced Technology Center, Org. A021S, Bldg. 252, 3251 Hanover St., Palo Alto, CA 94304, USA}

\date{Received 2 February 2018 / Accepted 7 March 2018}

\abstract{Low-lying loops have been discovered at the solar limb in transition region temperatures by the \emph{Interface Region Imaging Spectrograph} (IRIS). They do not appear to reach coronal temperatures, and it has been suggested that they are the long-predicted unresolved fine structures (UFS). These loops are dynamic and  believed to be visible during both heating and cooling phases. Making use of coordinated observations between IRIS and the Swedish 1-m Solar Telescope, we study how these loops impact the solar chromosphere. We show for the first time that there is indeed a chromospheric signal of these loops, seen mostly in the form of strong Doppler shifts and a conspicuous lack of chromospheric heating. In addition, we find that several instances have a inverse  Y-shaped jet just above the loop, suggesting that magnetic reconnection is driving these events. Our observations add several puzzling details to the current knowledge of these newly discovered structures; this new information must be considered in theoretical models.
}


\maketitle

\section{Introduction}

The solar transition region  (TR)  lies between the hot corona and the dynamic chromosphere. The origin and configuration of plasma between temperatures of a few $10^4$~K and $10^6$~K has long been debated. \cite{Feldman:1983} challenged the view that the chromosphere, TR, and corona connect in one continuous structure, and proposed the existence of `unresolved fine structure' (UFS) where smaller loops reach TR temperatures but do not connect to the corona. The existence of such small loops would help explain why the observed emission in TR lines is orders of magnitude higher than the predicted emission from models, already noted by \cite{Gabriel:1976}. It was only recently with the advent of the \emph{Interface Region Imaging Spectrograph} \citep[IRIS, ][]{IRIS-paper} that the necessary spatial resolution to resolve such loops was achieved.

\cite{Hansteen:2014} have recently discovered several small loops in IRIS observations (hereafter referred to as low-lying loops) and suggest that they are the long-postulated UFS. The loops have scales of about 2--6~Mm and are very dynamic, often showing high velocities as deduced from strong velocity shifts (80~$\kms$ or more).
Seen by IRIS, these loops are not as abundant as `spicules' (here we use the term for both  limb and disk objects), but they are nevertheless a unique laboratory for the study of small-scale energy release up to TR temperatures, and may shed light onto how such processes scale to larger structures and drive coronal heating. \cite{Brooks:2016} investigate whether these loops are really spatially resolved by IRIS, or if the heating is taking place in smaller, multiple magnetic threads. The authors measured the properties of many of these loops and compared them with a hydrodynamic simulation of an impulsively heated single strand. They conclude that the observed spatial scales, lifetimes, and heating profiles are consistent with single-strand heating and that UFS low-lying loops may be resolved by IRIS.

Typical coronal loops are observed mostly in their cooling phase \citep{Ugarte-Urra:2009}, but  both heating and cooling phases can be observed for low-lying loops  \citep[they light up in segments, see][]{Hansteen:2014}. This makes it possible to study the onset of heating in detail. In this work we look into the chromospheric response of these low-lying loops in an attempt to better understand their formation and relation to surrounding layers. We report for the first time a chromospheric signal associated with these loops, seen not in intensity enhancements but in strong velocities, and present evidence that magnetic reconnection may be at play in some of the loops.

\begin{figure*}
\begin{center}
\includegraphics[scale=1]{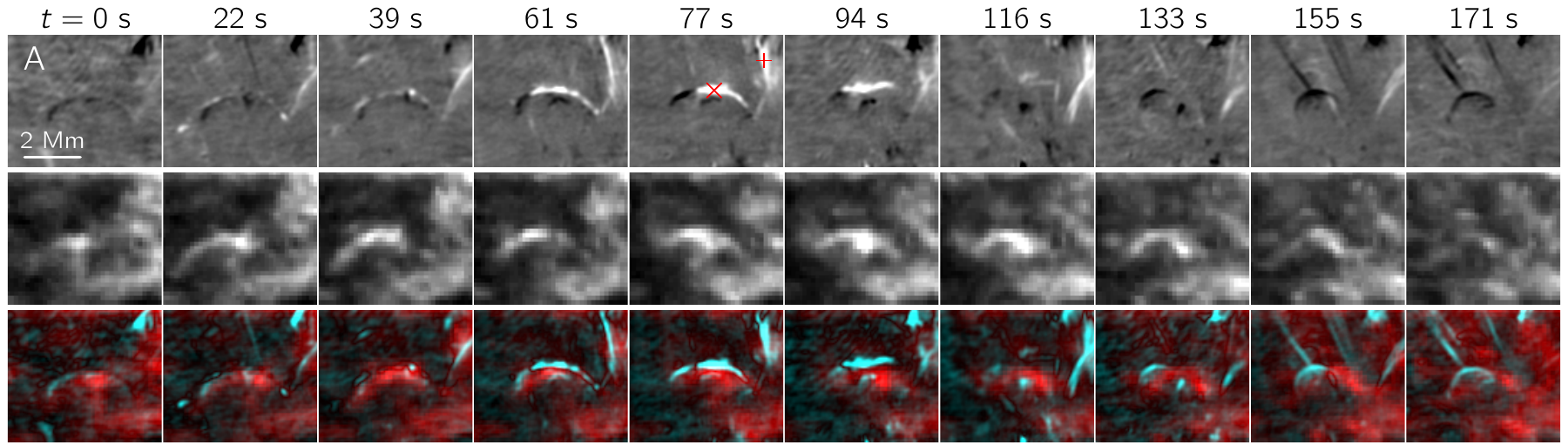}
\includegraphics[scale=1]{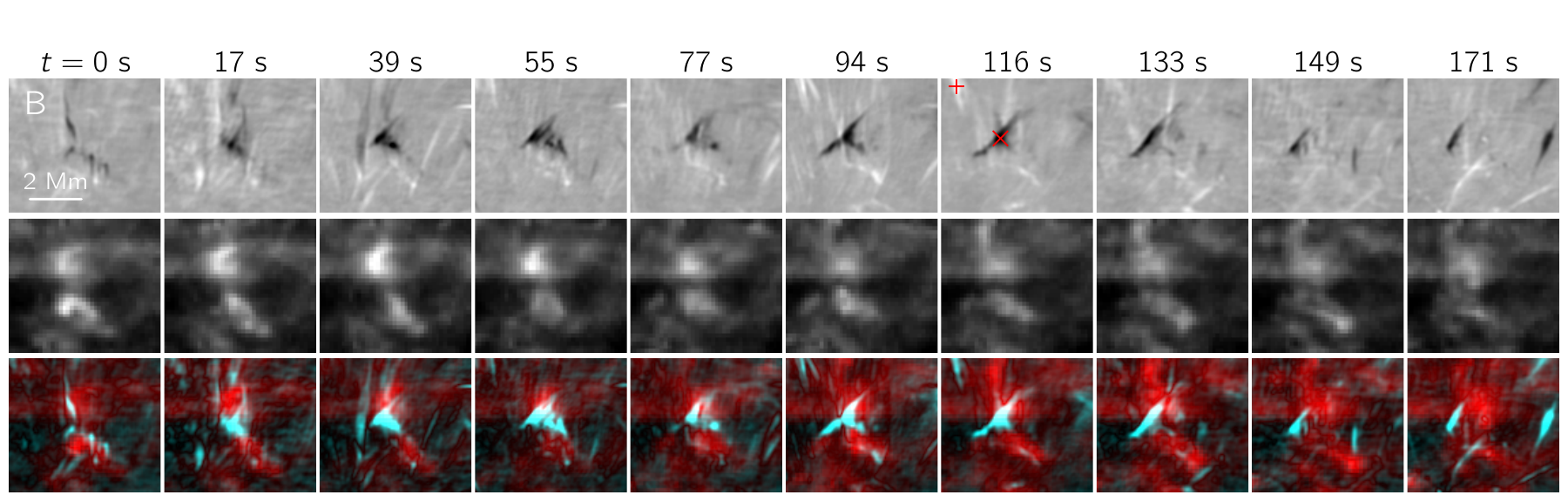}
\end{center}
\caption{Evolution of low-lying loops. Two events A and B are shown  (top and bottom panel, respectively). The different rows in each panel show \ha\ $\pm 55\,\kms$ Dopplergrams (top, white/black represent blueshifts/redshifts), IRIS SJI 1400 (middle, closest in time to the SST), and a combined image  (bottom, red for SJI 1400 and cyan for the absolute value of the \ha\ Dopplergrams). For each event, one of the \ha\ frames shows a red cross and plus sign, which  are the chosen locations for the spectra shown in Figure~\ref{fig:spectra}. (An animation of this figure is available online on arXiv under `Ancillary files'.)
\label{fig:evolution}}
\end{figure*}

\section{Observations}

We obtained coordinated observations between the Swedish 1-m Solar Telescope \citep[SST,][]{Scharmer:2003} and IRIS. We make use of the dataset from 17 June 2014 described by \citet{Pereira:2016} (to which we refer  for more details about the SST data and its reduction), and a similar dataset acquired two days later. Data were obtained at the SST with the Crisp Imaging SpectroPolarimeter \citep[CRISP, ][]{Scharmer:2008}. With CRISP we scanned \ha\ along 25 positions from $-0.12$ to 0.12~nm around the line core in 0.01~nm steps. The cadence of a full scan was 5.5~s and the field of view was approximately $61\arcsec\times61\arcsec$, with a plate scale of $0\farcs058$~pixel$^{-1}$. For the first dataset (17 June) we also acquired \CaH\ filtergrams using an interference filter with a FWHM of 0.11~nm centred at 396.88~nm.
IRIS was running a `sit and stare' programme (OBS-ID 3820259453) with the slit tracking a fixed location on the solar  surface, and included the slit-jaw filters 140.0~nm (SJI 1400) and 279.6~nm (SJI 2796) . We use level 2 calibrated data, mainly from the SJI 1400 images, whose cadence was 18.8~s. These filtergrams are dominated by \ion{Si}{iv} emission, sensitive to temperatures around $65\,000$~K \citep{IRIS-paper}.

The target of the observations was quiet Sun at the solar north pole. For June 17 we used data observed between 10:20 UT and 11:15 UT, with the SST field of view centred  at solar $(x, y)$ coordinates of (24$\arcsec$, 939$\arcsec$), while on June 19 the data were taken between 08:30 UT and 09:46 UT and the centre was (13$\arcsec$, 909$\arcsec$), slightly closer to disk centre. There was no polar coronal hole during either period. The CRISP data were reduced using the CRISPRED pipeline \citep{de-la-Cruz-Rodriguez:2014}, and we made use of the Multi-Object, Multi-Frame Blind Deconvolution image restoration technique of \citet{van-Noort:2005}, and employed the method of \citet{Henriques:2012} to minimise residual seeing deformations.

For June 17, a precise co-alignment between SST and IRIS was achieved by using the SST \CaH\ filtergrams, which were internally co-aligned with the CRISP \ha\ images and later with the IRIS SJI 2796 (\MgIIk) images. For June 19, no SST \CaH\ images were available; we used AIA \citep{Lemen:2012} and HMI \citep{Schou:2012} as an intermediate step to align the IRIS and  CRISP images. The HMI continuum was aligned with the \ha\ wideband and the AIA 160~nm channel (previously co-aligned with HMI continuum) was aligned with the IRIS SJI 1400 images. We estimate the accuracy of the June 17 co-alignment to be around $0\farcs 2$, and slightly worse for June 19.

Low-lying TR loops were visually identified using the IRIS SJI 1400 images. We chose events where a loop or part of a loop was clearly visible above the background, and chose mostly events on disk. At the limb there is a lot of superposition with spicules, making it hard to find the \ha\ counterpart of the loops. We include only one example at the limb, a particularly strong and clear event.
Low-lying loops clearly visible in SJI 1400 on the solar disk are relatively rare.
In about 2~h of observations we found 14 examples. \citet{Brooks:2016} found 108 loops in 5~h, with a field of view about 4 times larger (our analysis was restricted to CRISP's $61\arcsec\times61\arcsec$ field of view), so the frequency of loops is comparable.

\section{Results}

We find that most TR loops do not show up prominently in \ha\ narrow-band images, except in the far wings. They are more clearly seen as absorption features in the blue or red wings of \ha. Subtracting two narrow-band images at a symmetric distance from the line core we construct Dopplergrams (here shown at $\pm 55\,\kms$), useful to reveal events with strongly Doppler shifted \ha\ wings, such as spicules and also low-lying loops.

We show the time evolution of two events in Figure~\ref{fig:evolution} and the accompanying movies. Both events have half-lengths of about $1.6-2$~Mm, consistent with the typical scales found by \citet{Brooks:2016}. We show \ha\ Dopplergrams, SJI 1400 images and composite images to better visualise the spatial overlap of the two. Event A is more clearly visible as a loop in both \ha\ and SJI 1400 images. One can see a loop shape in \ha\ Dopplergrams even before it is clearly defined in IRIS. A complex picture of blueshifts and redshifts is visible in \ha: it starts as the whole loop slightly redshifted, but turns to strong blueshifts coming from the right footpoint and dominating the whole loop. At $t\approx94$~s there is a powerful ejection and what is seen as the right footpoint is not only blueshifted, but also ejected upwards.
This `ejection phase' coincides with an increase in the loop brightness in SJI 1400, near the right footpoint. After the ejection, the loop is still clearly visible in SJI 1400 but disappears from \ha, only showing up as a smaller redshifted loop at $t\approx133$~s that may not necessarily be related to the first event.

\begin{figure}
\begin{center}
\includegraphics[scale=1]{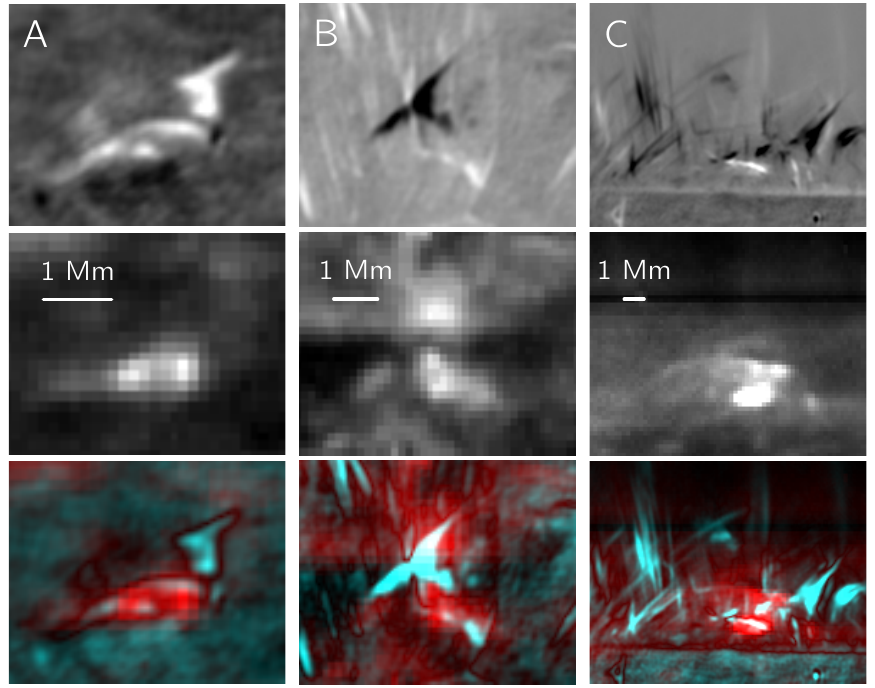}
\end{center}
\caption{\ha\ inverse Y-shapes and associated SJI 1400 loops for three events (A, B, C). As in Figure~\ref{fig:evolution}, the different rows show \ha\ $\pm 55\,\kms$ Dopplergrams (top), IRIS SJI 1400 (middle), and a combined image  (bottom, red for SJI 1400 and cyan for the absolute value of the \ha\ Dopplergrams). Each panel has been individually scaled for the best contrast.
\label{fig:reversey}}
\end{figure}

Event B has a more diffuse shape in SJI 1400 images and does not always appear as a loop. In SJI 1400 it also has a brighter structure immediately above that seems unrelated to the event. The absence of a full loop is consistent with earlier findings \citep{Hansteen:2014, Brooks:2016} and could be related to the episodic heating suggested by both sets of authors. In \ha\ the region is crowded with spicules (white, redshifted features) and the loop is even less clearly a loop. What is seen instead is a dark, redshifted structure hovering over the SJI 1400 loop, with an inverted Y-shape similar to `anemone' jets \citep{Shibata:2007}. The overlap between the \ha\ structure and the SJI 1400 loop is more evident between $77<t<133$~s. At $t\approx94$~s the bottom of the anemone jet seems clearly above the SJI 1400 loop. Both the loop and the \ha\ structure disappear at around the same time ($\approx 200$~s after the event is visible). While this could be a chance alignment between two unrelated structures, the temporal and spatial coincidences are striking.

\begin{figure}
\begin{center}
\includegraphics[scale=1]{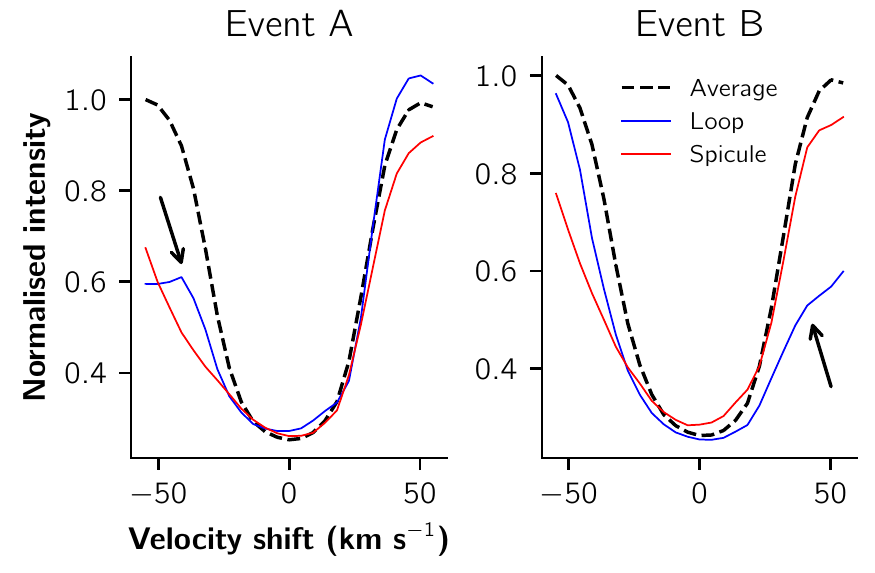}
\end{center}
\caption{\ha\ spectra of low-lying loops and spicules. Taken from events A and B from Figure~\ref{fig:evolution}, compared with the mean spectrum averaged over the field of view of the panels in Figure~\ref{fig:evolution} and the whole time sequence (black dashed line). The spectra of the low-lying loops (blue lines) and spicules (red lines)  were observed at the locations of the red crosses and plus signs, respectively. Arrows indicate knees in the profiles.
\label{fig:spectra}}
\end{figure}

Looking at other examples of low-lying loops, we see similar inverted Y-shaped \ha\ structures above or very close to the SJI 1400 loops. We show a larger image of the same event B and two other different occurrences in Figure~\ref{fig:reversey}. In event A in Figure~\ref{fig:reversey} (which is not the same event in Figure~\ref{fig:evolution}) we see another hint of an anemone jet in \ha\ just above a brightening in SJI 1400 that is also seen as a loop shape in \ha. Event C was observed above the limb (the limb is visible in the bottom part of the image). Its field is crowded with spicules, the SJI 1400 loop is larger than the previous examples (about 5 Mm half-length) and shows a strong brightening just below the loop. In event C there is an anemone jet in the vicinity of the large SJI 1400 loop (which appears  dark on the right side of the Dopplergram), but co-spatiality is harder to establish. It could be an unrelated, superimposed structure. However, we include it to show a pattern of what might or might not  be coincidences, and also to show an example above the limb.

In our \ha\ Dopplergrams  spicules are often present close to the low-lying loops. Spicules are ubiquitous in such images \citep[see e.g.][]{Pereira:2016}, and the regions around the loops are no exception. Morphologically, it is easy to distinguish the two:  spicules are straighter and more elongated. Spectrally, the distinction is not as clear because  both show red or blueshifted \ha\ wings, depending on the geometry and dominant motion \citep{DePontieu:2012}. From our June 17 dataset we note that at some point nearly all loops show  extreme Doppler shifts and a peculiar spectral profile that is seldom, if ever, seen in even the strongest spicules. Some examples are shown in Figure~\ref{fig:spectra}, taken from the events of Figure~\ref{fig:evolution}. The two cases show strong Doppler shifts in both loops and spicules. We chose the strongest spicule in each field of view. The spectra of the two loops (for completeness we show both red and blueshifted events) show strong Doppler shifts that have a severely depressed line wing. Although limited by our spectral coverage (from $-55$ to $55\;\kms$ around \ha) the loop spectra display a knee in the wing reminiscent of a strongly Doppler shifted component of chromospheric plasma (see arrows in Figure~\ref{fig:spectra}).
It is different from the typical spicule profiles that show wing absorption but not usually a strong knee  \citep[see e.g.][]{vdVoort:2009}. However, such extreme spectra are absent from the loops in our June 19 dataset (closer to disk-centre), suggesting that a favourable inclination is needed to see such high line of sight velocities.

A chromospheric signal of the SJI 1400 loops is not seen in our \CaH\ filtergrams \citep[not even footpoint brightening, as suggested by][in their anemone Ca jets]{Shibata:2007}, in IRIS SJI 2796, or in \ha\ close to line core possibly because they are obscured by fibrils which become opaque at the core of chromospheric lines. Of the 14 loops we found  in SJI 1400, only two do not have a clear signal in \ha. Conversely, we also found a few similar rounded loops in \ha\ that do not show any SJI 1400 brightenings. However, most of these \ha\ counter examples have larger spatial scales than the typical low-lying loop, and many seem related to curved spicules.

\section{Discussion}

We set out to find the chromospheric response of low-lying transition region loops observed by IRIS, hoping to find clues to their heating and formation process. We found that the loops appear mostly as strong Doppler shifts that either trace the IRIS SJI 1400 loops, or appear just above the loop itself.
Nearly all loops have footpoints that seem either shared with or in very close proximity to spicule footpoints, predominantly rooted in the magnetic network. Unlike spicules, the loops do not exhibit \CaH\ intensity brightenings or enhanced \ha\ widths \citep[see][for their relation to temperature]{Leenaarts:2012halpha}, suggesting that most loop material is at TR temperatures. The line of sight and apparent motions observed in \ha\ show that violent motions of cool plasma can occur near and above the loops. Furthermore, the appearance of inverse Y-shaped structures above the loops suggests that magnetic reconnection is taking place. However, it should be noted that our observations show no trace of footpoint brightening and the jets are not as long as those reported by \citet{Shibata:2007}. Yet several details are puzzling: if the inverse Y-shaped structures are indeed caused by magnetic reconnection, we would expect strong heating and therefore intensity brightenings, but the inverse Y is seen only in the far wings of \ha, not in other chromospheric or TR filtergrams. Why would the TR heating be limited to the `loop' part of the structure?

The observations show a definite connection between the TR loops and the chromosphere, but the processes by which they are related are unclear.
Based on our observations, it is clear that the events are more complex than a single loop, a scenario which fails to describe some of the puzzling findings.
We suggest another possible scenario: these structures are not loops but arcades (or domes) that emerge from below, pushing some of the cool chromospheric plasma into higher layers \citep[possibly related to the mechanism described by][]{Ortiz:2016}. When a trigger mechanism takes place, possibly magnetic reconnection, heating can still take place along a single or multiple strands with the characteristic spatial width of $\approx 133$~km as found by \citet{Brooks:2016}, showing up as a loop in TR images. Whatever the cause, it is clear that there are violent motions along the surface of the arcade, sometimes in opposing directions, as evidenced by the \ha\ observations. In some cases, such as event A in Figure~\ref{fig:evolution}, there is a strong ejection of the cool chromospheric material while the TR heating continues below. In other cases (e.g. where a inverse Y-shape is visible), the orientation of the motion is such that the cool plasma seems to trace the magnetic field lines. Depending on where in the loop reconnection takes place, we were able to  observe both upflowing and downflowing plasma, which can help explain the strongly shifted \ha\ spectra.

Our findings should be considered exploratory and not general properties of all low-lying loops in the quiet Sun. Our sample is limited to  14 loops. Nevertheless, we find evidence that nearly all TR loops have a chromospheric counterpart. Even if magnetic reconnection is not the main driver of the loops, it is clear that there are ejections from, or associated with, a subset of loops in chromospheric layers. They could be caused by the loop driver or by interaction with the surrounding magnetic field. These ejections have not been reported before, and add a new aspect to consider when modelling the evolution of the loops.

\section{Conclusions}

We show how low-lying TR loops (UFS loops), an intriguing class of objects recently discovered with IRIS, appear in chromospheric diagnostics. We find subtle traces of the loops in the chromosphere: no brightenings or evidence of heating, but strong velocities both apparent and in extreme Doppler shifts. It appears that cool chromospheric plasma exists around the hot TR loops and is accelerated to high speeds. The driver of both TR heating and chromospheric motions is unclear, but we find evidence that magnetic reconnection may be taking place. In about one-third of the detected loops we find a  inverse Y-shaped structure above or near the TR loops in \ha\ wing images, suggesting reconnection according to the the anemone jet scenario of \citet{Shibata:2007}. We find that low-lying loops have footpoints that are very close to those of spicules and can have similar red and blueshifted \ha\ spectra, but in some cases (or viewing angles) can show more extremely Doppler-shifted spectra than spicules.

If magnetic reconnection is the driver of these loops, they are among the smallest known ($< 2$~Mm) manifestations of reconnection in the quiet Sun.
However, our findings leave many questions unanswered and have several puzzling details. Why is the reconnection signature seen in cooler plasma above the hotter part of the loop? What keeps the chromospheric plasma in those regions, and why is it violently ejected but not heated? We hope that this exploratory work can be used to further constrain  the theoretical modelling of these objects and help explain how energy is channelled through the solar atmosphere.

\begin{acknowledgements}
We would like to thank Rob Rutten for the alignment of SST and AIA data for the June 19 dataset. This work was supported by the European Research Council under the European Union's Seventh Framework Programme (FP7/2007-2013) / ERC Grant  agreement No. 291058,
and by the Research Council of Norway, project number 250810, and through its
Centres of Excellence scheme, project number 262622. B.D.P. gratefully acknowledges support from NASA grants NNX16AG90G and NNG09FA40C (IRIS). IRIS is a NASA small explorer mission developed and operated by LMSAL with mission operations executed at NASA Ames Research Center and major contributions to downlink communications funded by ESA and the Norwegian Space Centre. The Swedish 1-m Solar Telescope is operated on the island of La Palma by the Institute for Solar Physics of Stockholm University in the Spanish Observatorio del Roque de los Muchachos of the Instituto de Astrof\'\i{}sica de Canarias. To visualise, connect, and interpret the data we made extensive use of CRISPEX \citep{Vissers:2012}.
\end{acknowledgements}

\bibliographystyle{aa}

\end{document}